# Controlling the quantum dynamics of a mesoscopic spin bath in diamond

Gijs de Lange[1†], Toeno van der Sar[1], Machiel Blok[1], Zhi-Hui Wang[2], Viatcheslav Dobrovitski[2], and Ronald Hanson[1*]

*1. Kavli Institute of Nanoscience Delft, Delft University of Technology,*

*P.O. Box 5046, 2600 GA Delft, The Netherlands*

*2. Ames Laboratory and Iowa State University, Ames, Iowa 50011, USA*

[†]*e-mail: g.delange@tudelft.nl*

[*]*e-mail: r.hanson@tudelft.nl*

**Abstract**

**Understanding and mitigating decoherence is a key challenge for quantum science and technology. The main source of decoherence for solid-state spin systems is the uncontrolled spin bath environment. Here, we demonstrate quantum control of a mesoscopic spin bath in diamond at room temperature that is composed of electron spins of substitutional nitrogen impurities. The resulting spin bath dynamics are probed using a single nitrogen-vacancy (NV) centre electron spin as a magnetic field sensor. We exploit the spin bath control to dynamically suppress dephasing of the NV spin by the spin bath. Furthermore, by combining spin bath control with dynamical decoupling, we directly measure the coherence and temporal correlations of different groups of bath spins. These results uncover a new arena for fundamental studies on decoherence and enable novel avenues for spin-based magnetometry and quantum information processing.**

In the past few years, new advances in quantum science and technology have underscored the importance of understanding and controlling decoherence of single solid-state spins[1-3]. Decoherence of a single central spin in contact with a spin bath environment has been intensively studied in various systems such as quantum dots[4-7], donors in silicon[8] and defects in diamond[9-12] through control and readout of the central spin. Here, we implement quantum control of both the central spin and its spin bath environment, thereby enabling a range of new experiments on the fundamentals of decoherence. Moreover, spin bath control is a crucial ingredient of recent proposals for environment-assisted magnetometry[13], room-temperature quantum computing using spins in diamond[14,15] and spin squeezing[16].

Our study focuses on the electronic spin bath environment formed by nitrogen impurities surrounding a single NV centre in diamond (Fig 1a). The electron spin of the NV centre can be initialized and read out optically, and coherently controlled with high fidelity at room temperature using microwave magnetic pulses[17-19]. For controlling the quantum state of the bath spins, we apply short (tens-of-nanoseconds) radiofrequency (RF) pulses to the sample. The high intensity control fields that allow fast control of both the central NV spin and the bath spins are delivered through a broadband coplanar waveguide (CPW) fabricated on the diamond substrate.

The state of the (optically inactive) spin bath can in principle be monitored directly via the emitted RF radiation as in conventional electron spin resonance. However, this method requires many orders of magnitude more spins than contained in our mesoscopic region of interest, and is limited to high magnetic fields. Instead, we exploit the coupling of the bath spins to the single NV centre. The near-atomic size of the NV centre, combined with the strong ($\sim 1/r^3$) distance dependence of the dipolar coupling to the surrounding bath spins, renders the NV spin mainly sensitive to a small number $N$ (a few tens) of bath spins. This local spin bath exhibits a large statistical polarization ($\sim 1/\sqrt{N}$) that is felt by the NV centre as a magnetic dipolar field $\delta b$. The spin bath polarization and the corresponding value of the bath field $\delta b$ change in time due to flip-flop processes within the bath, leading to dephasing of the NV centre spin on a timescale $T_{2,NV}^*$ of about 300 ns[19]. This quasi-static dephasing is compensated in a spin echo sequence with a refocusing π-pulse (Fig.1b), yielding decay on a much longer timescale $T_{2,NV} = (2.6 \pm 0.1)$ μs[19]. However, if we induce changes in the state of the bath spins (thus changing the value of $\delta b$) by applying a high intensity RF pulse halfway the NV spin echo sequence, the refocusing is ineffective and the NV spin echo amplitude is reduced (Fig.1b). Therefore, by incorporating the spin bath control within a spin echo sequence of the NV centre the resulting spin bath dynamics can be probed.

**Results**

**Magnetic resonance spectroscopy of bath spins.** To identify the environmental spins we perform magnetic resonance spectroscopy by sweeping the frequency of applied RF pulses while monitoring the NV spin echo amplitude (upper panel of Fig 1c). Several sharp dips are observed, demonstrating that spins in the environment are being rotated at these specific frequencies. The obtained spectrum matches that of single electron spins ($S = 1/2$) belonging to substitutional nitrogen (N) impurities[20]. We find excellent agreement with a theoretical spectrum (lower panel Fig. 1c) calculated using known values for the Zeeman energy and hyperfine interaction with the N nuclear spin, which is anisotropic due to a static Jahn-Teller distortion (Fig. 1a). Since the resonance frequencies are spaced by several line widths, only spins that belong to the same spectral group can exchange energy via flip-flop processes. The spin environment of the NV centre can therefore be decomposed into different spectral groups of electron spins (labelled I to V) that are distinguished by their hyperfine interaction with the host N nuclear spin.

With the resonance frequencies known, we can coherently control the spin environment. Fig. 2a shows the effect of short RF pulses at the resonance frequency of group II spins. Periodic revivals in the NV spin echo amplitude are observed as a function of RF pulse length, with a frequency that increases with RF pulse amplitude. This behaviour is the key signature of coherently driven ("Rabi") oscillations, demonstrating that we have achieved quantum control of the spin environment. We note that the NV spin echo revives almost completely whenever the bath spins are rotated by a multiple of $2\pi$, indicating that the environment has returned to the state it had before the RF pulse. We can control all other spin bath groups in a similar manner (Fig. 2b). In addition, our setup allows us to rotate several or all of the groups simultaneously with multi-frequency control pulses.

**Spin echo double resonance (SEDOR).** The ability to control both the NV centre spin and its spin bath environment opens up a range of new possible experiments aimed at studying and manipulating the coupling between a central spin and a spin bath as well as investigating the internal bath dynamics. We first apply the bath control to measure the coupling of each of the bath spin groups to the NV centre spin using a spin echo double resonance (SEDOR) scheme[21] (Fig. 3a). With this scheme the dephasing of the central spin induced by one particular group of bath spins can be probed, while the effect of all other dephasing channels (including other spin bath groups) is refocused. We find that, whereas the NV spin echo amplitude decays as $\exp[-(t/T_{2,\text{NV}})^3]$, the SEDOR scheme yields a faster, Gaussian-shaped decay (see Fig.3a). The Gaussian shape indicates that the decay observed with SEDOR is dominated by the quasi-static static dephasing channel that we have selectively turned on. Therefore, the SEDOR decay time $T_{\text{SEDOR},i}$ directly yields the r.m.s. interaction strength $b_i$ between the NV spin and the $i^{\text{th}}$ spin bath group via $1/T_{\text{SEDOR},i} = b_i/\sqrt{2}$. We find $b_{\text{I}} = (0.83\pm0.02)$ µs$^{-1}$, $b_{\text{II}} = (1.59\pm0.03)$ µs$^{-1}$, $b_{\text{III}} = (1.58\pm0.04)$ µs$^{-1}$, $b_{\text{IV}} = (1.63\pm0.04)$ µs$^{-1}$ and $b_{\text{V}} = (0.80\pm0.02)$ µs$^{-1}$ (see Supplementary Fig. 1). These values are close to the ratio of $b_{\text{I}}:b_{\text{II}}:b_{\text{III}}:b_{\text{IV}}:b_{\text{V}} = 1:\sqrt{3}:2:\sqrt{3}:1$ expected from the abundance of each spectral group[20], except for the slightly lower value for group III. This group is actually composed of two subgroups which spectrally do not coincide perfectly. The control fidelity is therefore lower for this group which results in a lower measured coupling in the SEDOR experiment

The r.m.s. field fluctuations generated by the full electron spin bath are given by $b_{spin\ bath} = \sqrt{\sum_i b_i^2} = (3.01 \pm 0.04)$ μs$^{-1}$. This value falls short of the measured total dephasing rate of $b_{total} = (3.6 \pm 0.1)$ μs$^{-1}$ (see Supplementary Fig. 2a), suggesting the presence of additional dephasing channels, such as the carbon-13 nuclear spins[9,10] and magnetic field drifts, with a strength of $b_{excess} = \sqrt{b_{total}^2 - b_{spin\ bath}^2} = (1.97 \pm 0.04)$ μs$^{-1}$. This interpretation is supported by independent measurements on an NV centre in a pure diamond sample with low nitrogen content under the same experimental conditions that yield $b_{excess} = (2.06 \pm 0.04)$ μs$^{-1}$ (see Supplementary Fig. 2b).

For applications in quantum information processing[1] and spin-based dc-magnetometry[2,3], suppressing dephasing is crucial. We now demonstrate that quantum control of the spin bath can be used to eliminate the effect of the spin bath on the free evolution dynamics of the NV centre spin (see Fig. 3b). By flipping all bath spins, the interaction between the NV centre and bath spins can be time-averaged to zero. This procedure is akin to dynamical decoupling as recently demonstrated on single NV centre spins[19,22,23], but has the advantage that no control pulses on the NV centre itself are required. We find that a refocusing π-pulse applied simultaneously to all bath spins (Fig. 3b) increases $T_{2,NV}^*$ up to the limit set by $b_{excess}$, indicating that dephasing by the electron spin bath is suppressed. Similar enhancement is achieved by continuous driving of all bath spins (see Supplementary Fig. 2a).

**Coherence and temporal correlations of bath spins.** By combining the spin bath control with the ability to freeze the evolution of the NV spin by dynamical decoupling[19,22,23], coherence and temporal correlations during free evolution within the spin bath can be directly probed. We replace the single refocusing pulse of the spin echo sequence on the NV spin by a dynamical decoupling (DD) sequence with a net π-rotation (see Supplementary information). The DD sequence provides a means to temporarily turn off our sensor (the NV centre) as it is made insensitive to the magnetic environment for the duration of the DD sequence; the net π-rotation ensures that the refocusing action of the sequence is preserved. The two periods of free evolution $\tau_s$ of the NV spin now serve as sensing stages which each sample the dipolar field generated by the bath spins. The NV echo amplitude is therefore a measure of the correlation between the dipolar fields measured during the two sensing stages. While the sensor is switched off, we can apply multi-pulse RF sequences to individual spectral groups of bath spins to study their coherence during free evolution. An RF Ramsey sequence (Fig. 4a) and Hahn-echo sequence (Fig. 4b) is applied to spectral group $i$ to measure its spin dephasing time $T_{2,i}^*$ and coherence time $T_{2,i}$ respectively. Data is shown for spectral groups I and II.

The values we find for $T_{2,i}^*$ are similar for the two groups as expected, since all bath spins suffer from the same dephasing channels formed by spins from all groups.

From the value of $T_{2,i}^*$ we estimate the local density of bath spins to be about 100 parts per million (see Supplementary information).

The bath spin-echo sequence yields different decay times, $T_{2,I} = (1.9 \pm 0.6)$ μs and $T_{2,II} = (0.89 \pm 0.13)$ μs, for spectral groups I and II. The difference in coherence times between different spectral groups may arise due to dephasing caused spins within the same group, in a process which is known as instantaneous diffusion[24,25]. The RF π-pulse does not refocus the dipolar interactions between spins of the same spectral group since these spins are themselves rotated by the RF π-pulse. The resulting intra-group dephasing is much stronger in group II because it contains three times more spins than group I.

To characterize the temporal correlations resulting from the dynamics in the environment we perform a direct measurement of the auto-correlation function and its $1/e$ decay time $\tau_C$ (see Supplementary information). The field generated by the complete magnetic environment is sampled during two sensing stages separated by a variable waiting time during which we turn off the NV centre sensor (Fig. 4c) and let the spin bath evolve freely[26]. As the sensor off-time is increased, the initial correlation between the two fields is gradually lost resulting in decreasing NV echo amplitude. We observe a decay of the auto-correlation function on a timescale of about 20 μs.

We can also find the correlation time of an individual spin bath group by inserting a SEDOR sequence in the sensing stage, as demonstrated for group II (Fig. 4c). The measured correlation time of group II is comparable to that of the complete magnetic environment, indicating that the coherence time of the NV centre is indeed limited by the dynamics of the electron spin bath. The measured correlation time is comparable to the value $\tau_c = b^2_{\text{spin bath}} T^3_{2,NV}/12 \approx 13$ μs expected from mean-field theory[19,24].

**Discussion**

In conclusion we have demonstrated full quantum control of a spin bath surrounding a single NV centre. These results pave the way for a new class of experiments on spin bath decoherence, such as manipulating the correlation time of different spin bath groups and generating squeezed spin bath states[16]. Furthermore, the suppression of spin dephasing by spin-bath control may be exploited for protecting coherence in spin-based quantum technologies[1,2,3,14,27,28]. Finally, quantum control of nitrogen electron spins close to NV centres as demonstrated here enables implementation of quantum registers of individual N electron spins[29,30], scalable coupling of NV centre quantum bits via spin chains[15] and ultra-sensitive environment-assisted magnetometry[13].

**Methods**

**Device and setup.** The diamond sample is a single crystal type Ib plate from Element Six, with a Nitrogen concentration specified to be below 200 parts-per-million. A high-bandwidth golden coplanar waveguide (CPW) structure for magnetic resonance is fabricated by electron beam lithography directly on top of the bulk diamond sample. A static magnetic field of 132 G aligned along the symmetry axis of the NV centre is generated by a custom-built four-coil vector magnet setup (Alpha Magnetics). Spin-dependent fluorescence of the NV-centre is detected using a home-built confocal microscope. To determine that the detected fluorescence originated from a single NV centre, we performed a measurement of the second-order intensity autocorrelation function $g^{(2)}(\tau)$ to verify that the antibunching dip reached a value well below 0.5 $\left(g^{(2)}(0) = 0.15 \pm 0.1\right)$. Spin initialization and readout is achieved by 600 ns laser (532 nm) pulses. The timings of all microwave, radiofrequency and laser pulses are controlled by a 4-channel arbitrary waveform generator (Tektronix AWG5014). Each experiment is repeated for ~2M times to achieve statistical noise levels of order 1%. All experiments are performed at room temperature.

Phase and frequency control of MW and RF pulses is provided by IQ modulation using two vector sources (R&S SMBV 100A) with carrier set to 2.5 GHz and 400 MHz to control the NV spin and bath spins respectively. I and Q inputs of each vector source are controlled by two analogue channels of the AWG. Single and multi-frequency pulses on the RF channel are generated by proper IQ mixing to achieve image-frequency rejection. The modulated MW and RF signals are fed to two high power amplifiers (Amplifier Research 25S1G4 and 30W1000B respectively). After amplification the MW and RF signals are combined and delivered to the on-chip CPW. RF field inhomogeneities are <1 % over the relevant length scales (10 nm) and therefore do not limit the control fidelity of the bath pulses.

**Spin bath description**. Substitutional nitrogen impurities in diamond (also called P1 centres in literature) exhibit trigonal symmetry due to a static Jahn-Teller distortion in which one of the four N-C bonds is elongated. Most of the electron density is concentrated at the antibonding orbital belonging to this elongated N-C bond[31]. The close proximity of and partial overlap between the electron wavefunction and the N nucleus, combined with the trigonal symmetry, gives rise to a strong anisotropic hyperfine interaction between the defect's electronic spin ($S = 1/2$) and the host N nuclear spin ($I = 1$). The internal Hamiltonian of a nitrogen impurity in diamond is given by

$$\hat{H}_{int} = A_{//}\hat{S}_z\hat{I}_z + A_{\perp}\left(\hat{S}_x\hat{I}_x + \hat{S}_y\hat{I}_y\right) - P\hat{I}_z^2 \qquad (1)$$

with $A_{//} = 114$ MHz, $A_{\perp} = 86$ MHz and $P = 4.2$ MHz and $\hat{S}$ and $\hat{I}$ are the operators for the electron and nuclear spin respectively of the nitrogen impurity[20]. The direction of the z-axis is set by the JT distortion axis and is therefore oriented along one of the four N-C bonds[20]. The outer two satellites (belonging to group I and V) in the spectrum of Fig. 1c result from nitrogen impurities which have a JT axis oriented along the externally applied field.

Both the reorientation of the JT axis[32] and the nuclear spin-flips of the nitrogen impurities occur on timescales ranging from milliseconds to seconds. This is much shorter than the time required to build up statistics (tens of minutes per data point) but much longer than the time it takes to perform a single experimental run (several microseconds). These processes can therefore be treated as being quasi-static within one run, while at the same time we are averaging in the complete experiment over the full thermal distribution of all nuclear spin projections and JT orientations. Also, the differences in local configurations of N impurities around different NV centres will change the $T_2^*$ and $T_2$ parameters of the NV spin, but not the relevant ensemble properties of the spin bath[33].

**Acknowledgements**

We sincerely thank D. D. Awschalom, S. Frolov, G. D. Fuchs, K. Nowack, D. Ristè and L. M. K. Vandersypen for useful discussions. We gratefully acknowledge support from the Defense Advanced Research Projects Agency, the Dutch Organization for Fundamental Research on Matter (FOM), the Netherlands Organization for Scientific Research (NWO), and the European Union SOLID programme. Work at Ames Laboratory was supported by the Department of Energy—Basic Energy Sciences under Contract No. DE-AC02-07CH11358.

**Author contributions**

G.d.L., T.v.d.S. and M.S.B. conducted the experiments, V.V.D. and Z.H.W. developed the theory. G.d.L. and R.H. wrote the paper. V.V.D. and R.H. supervised the project. All authors discussed the results, analysed the data and commented on the manuscript.

**Additional information**

**Supplementary Information** is available online

**Competing financial interests**

The authors declare no competing financial interests.

**Corresponding authors**

Correspondence to: Gijs de Lange or Ronald Hanson.

**Figure captions**

**Figure 1 | Magnetic resonance spectroscopy of a spin bath using a single spin sensor.
a,** Schematic of the system: a single NV centre electronic spin ($S = 1$) is surrounded by a bath of electron spins ($S = 1/2$) belonging to substitutional N impurities. The applied external magnetic field $B$ is aligned with the symmetry axis of the NV centre, which is oriented along the [111] crystallographic direction. Nitrogen impurities exhibit a static Jahn-Teller distortion, which results in an elongation of one of the four N-C bonds. As a result, the defect has a symmetry axis, also called the Jahn-Teller axis (indicated red), which is oriented along randomly along one of the crystallographic axes. There are therefore two geometric types of bath spins which are distinguished by their orientation of the Jahn-Teller axis relative to the external field $B$: those that have their JT axis at an angle $\alpha = 0^\circ$, and those with $\alpha = 109.5^\circ$. **b,** Measurement sequence for spin bath spectroscopy. A spin echo sequence is applied to the NV spin using MW pulses; the bath spins are controlled by RF pulses. The evolution of the NV spin during the sequence is sketched in the Bloch spheres at the bottom, both for the case of no spin bath control (solid line), and for the case of spin bath control applied (dashed line) (see Supplementary information for details). **c,** Upper panel: Magnetic resonance spectroscopy of the spin bath. A magnetic field of 132 G is applied along the NV centre symmetry axis. Roman numbers label the different groups of N electron bath spins, according to their nuclear spin projections $m_I$ and angle $\alpha$ between their Jahn-Teller axis and the external magnetic field: I,V: $m_I = \pm 1$ and $\alpha = 0°$, II,IV: $m_I = \pm 1$ and $\alpha = 109.5°$, III: $m_I = 0$ and $\alpha = 0°$ or $109.5°$. Lower panel: Calculation of the spectrum (see Supplementary information).

**Figure 2 | Coherent control of the spin bath. a,** Coherent driven oscillations of group II bath spins, using RF pulses of 298 MHz. Revivals in NV echo amplitude are observed whenever the bath spins from group II perform a $2\pi$ rotation. The maximum Rabi frequency extracted from fitting the upper trace at 20 dBm source power is $f_1 = (20.5 \pm 0.1)$ MHz. **b**, Independent coherent control of all groups of bath spins. Solid lines are fits to $\propto e^{-t/T_D} \cos(2\pi f_1 t)$ with $t$ the length of the RF pulse (see Supplementary information).

**Figure 3 | Control of NV spin coherence by spin bath manipulation. a,** Spin echo double resonance (SEDOR) experiment. The RF pulses have been calibrated to rotate a preselected group of bath spins over a $\pi$ angle. The NV spin echo curve is fit to $\propto \exp\left[-\left(2\tau/T_{2,\mathrm{NV}}\right)^3\right]$, SEDOR curves are fit to $\propto \exp\left[-\left(2\tau/T_{\mathrm{SEDOR},i}\right)^2\right]$ (see Supplementary information). **b**, Dynamical suppression of NV centre spin dephasing through spin bath control. Free evolution of the NV spin is shown with (blue) and without

(red) RF π-pulse applied to the bath spins. Solid lines are fits that include the detuning ($\Delta f = 30$ MHz) of the MW driving field compared to the NV spin splitting, and local hyperfine interaction of 2.2 MHz with the host nuclear N spin. This hyperfine interaction is responsible for the observed beating pattern (see Supplementary information). The overall signal decays with a Gaussian envelope, with decay constant $T^*_{2,NV} = (278 \pm 5)$ ns ($b_{\text{total}} = (3.60 \pm 0.06)$ μs$^{-1}$) in the absence of the spin bath control (red) and with decay constant $T^*_{2,NV} = (450 \pm 9)$ ns ($b_{\text{excess}} = (2.11 \pm 0.07)$ μs$^{-1}$) in the case where the bath control pulses are applied (blue).

**Figure 4 | Coherent dynamics and temporal correlations of the spin bath. a,** Measurement of decay during free evolution of spin bath groups I and II. The two sensing stages marked by $\tau_s$ serve to sample the magnetic environment before and after the Ramsey sequence is applied to the bath spins. The NV spin sensor is turned off while the RF Ramsey sequence is applied to the bath spins (see main text for details). Instead of detuning the RF pulse field with respect to the transition to observe fringes, an artificial detuning $f_a$ is introduced by changing the phase of the final RF π/2-pulse linearly with pulse separation $\varphi = 2\pi f_a \tau$. Solid lines are fits to

$$\propto y_0 + \sum_{i=\text{I,II}} A_i\, e^{-\tau/T^*_{2,i}} \cos\left[\left(2\pi f_a + \Delta_i\right)\tau + \phi_i\right]$$ (see Supplementary information for details).

Additional contributions resulting from the off-resonant driving of spins from the nearest neighbouring group is taken into account by an extra oscillating term with frequency $\Delta_i$, which is given by the detuning of this group with respect to the pulse carrier. The fast modulation of Ramsey fringes for group I (upper panel) result from off-resonant driving of the more abundant spins from group II ($\Delta_{II} = 2\pi \cdot (34.4 \pm 0.4)$ s$^{-1}$). From the fits we extract the decay constants $T_{2,I}^* = (97 \pm 11)$ ns and $T_{2,II}^* = (91 \pm 7)$ ns. **b,** Spin echo on spin bath groups I and II. The phase of the final RF π/2-pulse is changed as a function of total free evolution time $\tau$ as $\varphi = 2\pi f_a \tau$ with $f_a = 10$ MHz, resulting in oscillations in the NV spin echo amplitude with free evolution time $\tau$. Solid lines are fits to $\propto e^{-\tau/T_{2,i}} \cos(2\pi f_a \tau + \phi_i)$ from which the decay times $T_{2,I} = (1.9 \pm 0.6)$ μs and $T_{2,II} = (0.89 \pm 0.13)$ μs are extracted (see Supplementary information). **c,** Measurement of temporal correlations on the full environment and on spin bath group II alone. During the sensing stages marked by $\tau_s$, MW and RF π-pulses can be applied simultaneously to the NV spin and to the bath spins to selectively measure the correlation time of a particular group of bath spins. Solid lines are fits to $\propto \exp[-b^2 \tau_s^2 (1 - e^{-t/\tau_c})]$ (see Supplementary information).

Fig. 1

**a**

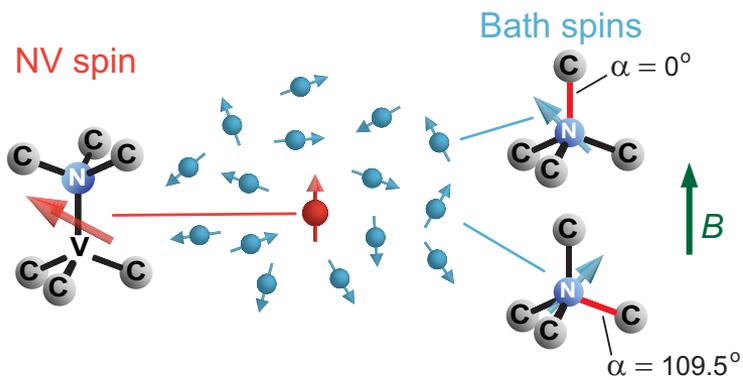

NV spin  Bath spins

$\alpha = 0°$

$B$

$\alpha = 109.5°$

**b**

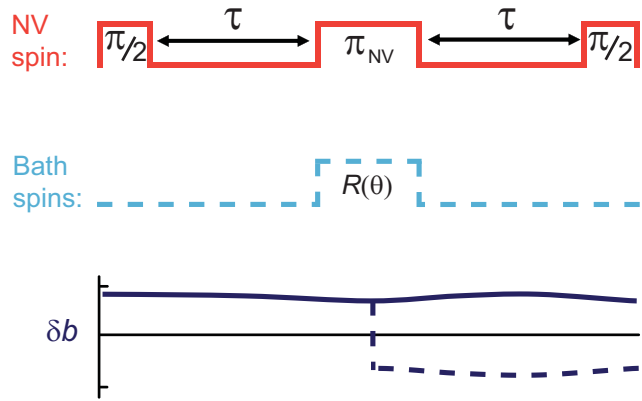

NV spin: $\pi/2$ — $\tau$ — $\pi_{NV}$ — $\tau$ — $\pi/2$

Bath spins: $R(\theta)$

$\delta b$

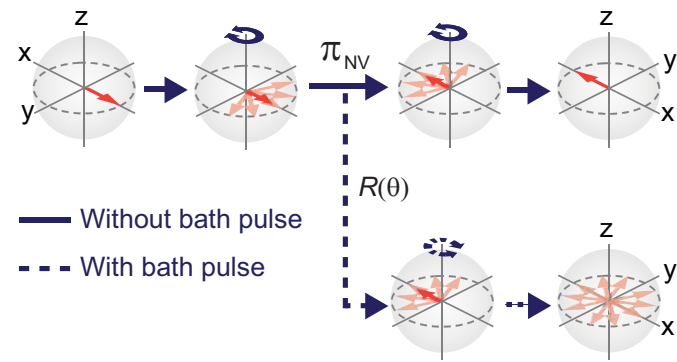

— Without bath pulse
-- With bath pulse

**c**

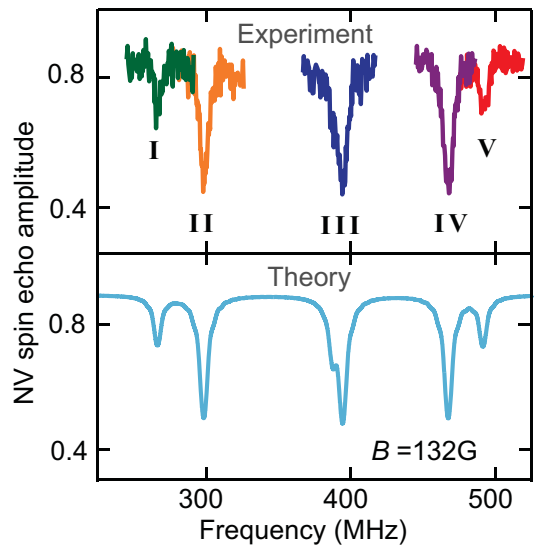

Experiment

I  II  III  IV  V

Theory

$B = 132\,G$

NV spin echo amplitude vs Frequency (MHz)

# Fig. 2

**a** 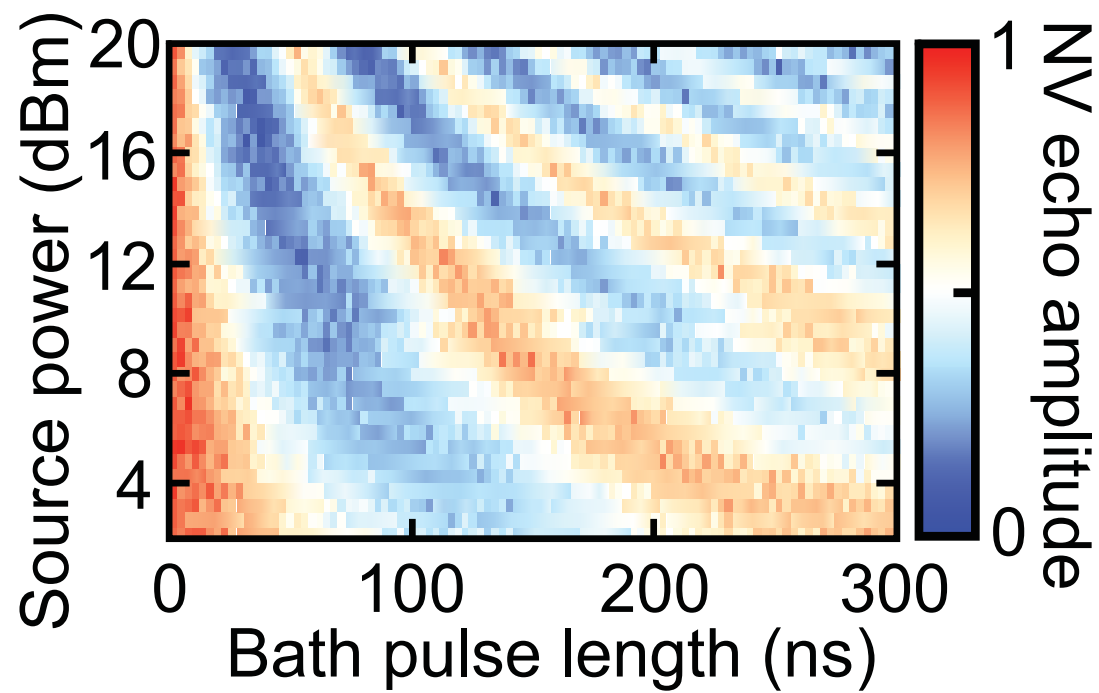

**b** 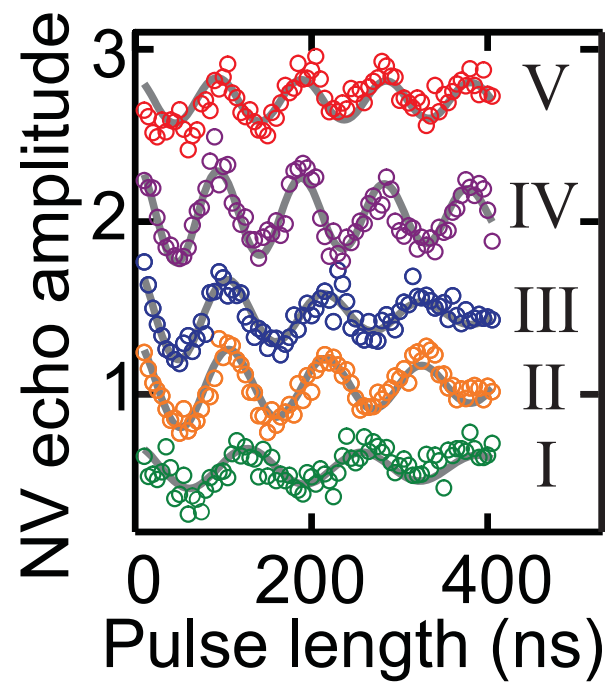

Fig. 3

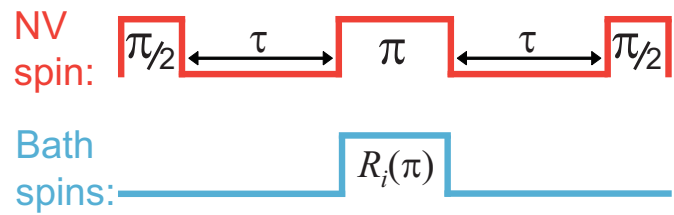
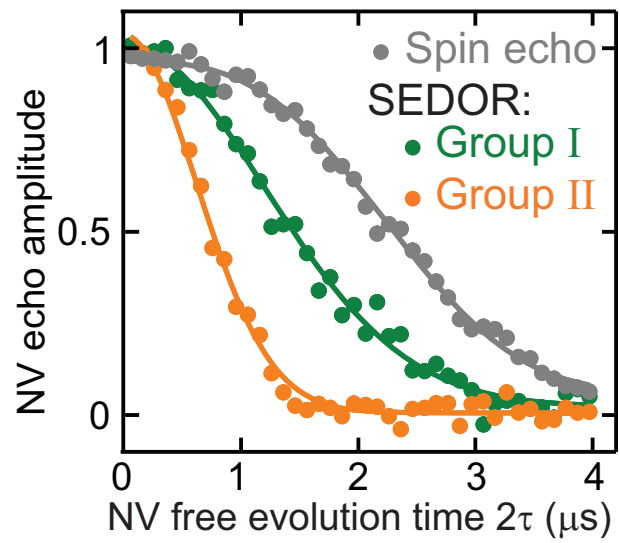

**a**

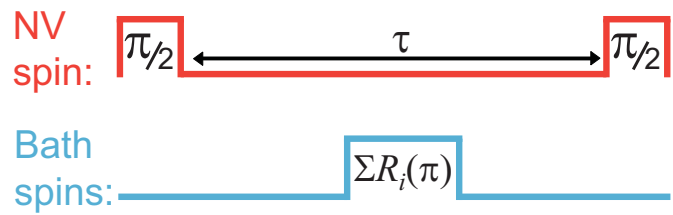
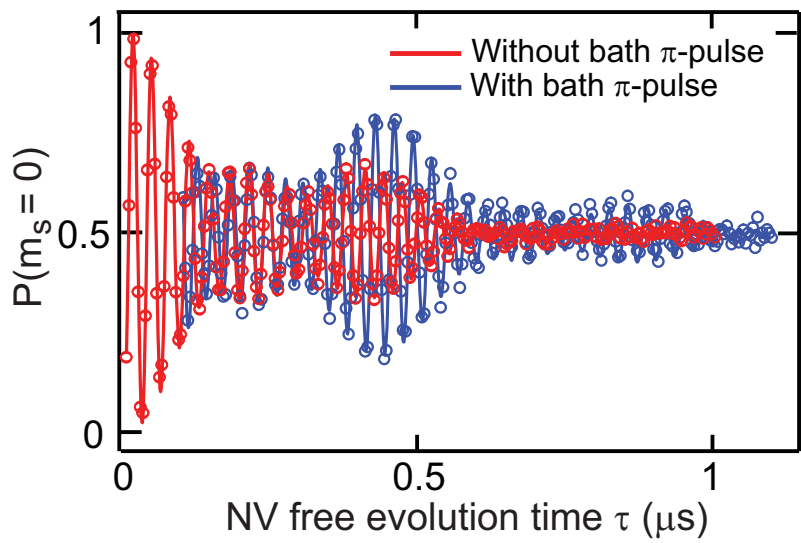

**b**

Fig. 4

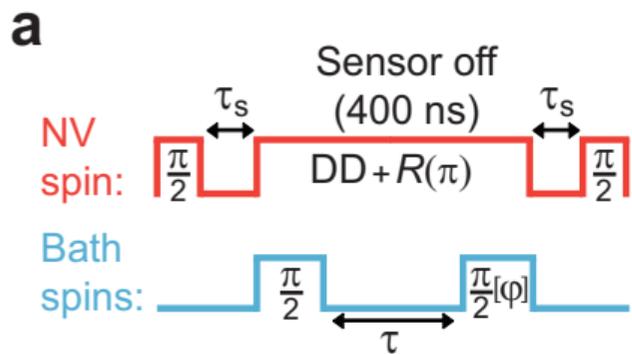 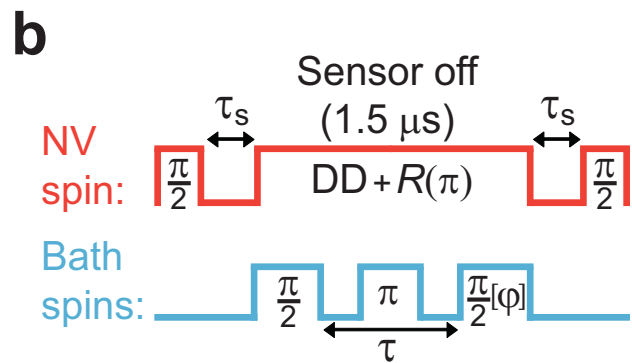 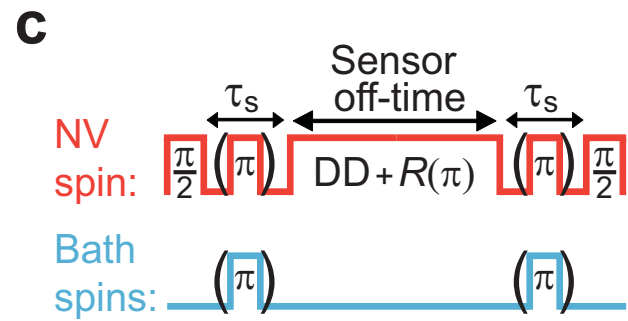